\documentclass{aastex}
\usepackage{emulateapj5}
\usepackage{apjfonts}

\newcommand{\ea}{{\it et al.}}

\newcommand{\jcomp}{\rmfamily{J. Comp. Phys.}}

\newcommand{\kms}{km~$\rm{s}^{-1}$}

\newcommand{\sfrac}[2]{\,{}^{#1}\!/_{#2}}
\newcommand{\beq}{\begin{equation}}
\newcommand{\eeq}{\end{equation}}
\newcommand{\bdm}{\begin{displaymath}}
\newcommand{\edm}{\end{displaymath}}

\slugcomment{Submitted to the Astrophysical Journal}

\begin{document}

\title{A Laboratory Investigation of Supersonic Clumpy Flows:
Experimental Design and Theoretical Analysis}

\author{A.Y. Poludnenko\altaffilmark{1,2}, K.K. Dannenberg\altaffilmark{3},
R.P. Drake\altaffilmark{3}, A. Frank\altaffilmark{1,2}, J. Knauer\altaffilmark{2},
D.D. Meyerhofer\altaffilmark{1,2,4}, M. Furnish\altaffilmark{5}, 
J.R. Asay\altaffilmark{6}}

\altaffiltext{1}{Department of Physics and Astronomy, University of Rochester, 
Rochester, NY 14627-0171; wma@pas.rochester.edu, afrank@pas.rochester.edu}
\altaffiltext{2}{Laboratory for Laser Energetics, University of Rochester,
250 East River Road, Rochester, NY 14623; jkna@lle.rochester.edu}
\altaffiltext{3}{Atmospheric, Oceanic, and Space Sciences, University of Michigan,
2455 Hayward Street, Ann Arbor, MI 48109; korbster@umich.edu, rpdrake@umich.edu}
\altaffiltext{4}{Department of Mechanical Engineering, University of Rochester, 
Rochester, NY 14627; ddm@lle.rochester.edu}
\altaffiltext{5}{MS 1168, Sandia National Laboratories, PO Box 5800, 
Albuquerque NM 87185-1168; mdfurni@sandia.gov}
\altaffiltext{6}{Washington State University, Institute for Shock Physics,
Box 642816, Pullman, WA 99164-2816; jrasay@wsu.edu}

\begin{abstract} 
  We present a design for high energy density laboratory experiments
  studying the interaction of hypersonic shocks with a large number of
  inhomogeneities. These ``clumpy'' flows are relevant to a wide
  variety of astrophysical environments including the evolution of
  molecular clouds, outflows from young stars, Planetary Nebulae and
  Active Galactic Nuclei. The experiment consists of a strong shock
  (driven by a pulsed power machine or a high intensity laser)
  impinging on a region of randomly placed plastic rods. We discuss
  the goals of the specific design and how they are met by specific
  choices of target components. An adaptive mesh refinement
  hydrodynamic code is used to analyze the design and establish a
  predictive baseline for the experiments.  The simulations confirm
  the effectiveness of the design in terms of articulating the
  differences between shocks propagating through smooth and clumpy
  environments. In particular, we find significant differences between
  the shock propagation speeds in a clumpy medium compared to a smooth
  one with the same average density. The simulation results are of
  general interest for foams in both inertial confinement fusion and
  laboratory astrophysics studies.  Our results highlight the danger
  of using average properties of inhomogeneous astrophysical
  environments when comparing timescales for critical processes such
  as shock crossing and gravitational collapse times.

\end{abstract}


\section{INTRODUCTION}

Advances in high-resolution imaging have revealed many astrophysical
environments to consist of highly inhomogeneous media. Those images
show that material on circumstellar, interstellar, and galactic scales
are not smooth plasma systems but are often arranged into a large
number of cloudlets or clumps immersed in a background of interclump
gas. The presence of such ``clumpy'' mass distributions may have
significant consequences for large-scale flows.  These flows can
dominate astrophysical processes.  Examples include: mass loss from
both young and evolved stars, strong shocks propagating through
interstellar clouds, and mass outflows from active galactic nuclei
(AGN). In each of these cases, a high velocity flow impinges on an
ambient medium that is accelerated, compressed, and heated.  The
momentum and energy exchange between the driver (the winds or
interstellar shocks) and the ambient medium can be an important source
of luminosity, non-thermal particles, mixing of enriched elements and
turbulence. Thus, the clumpy flows revealed in new images point to the
need for increased understanding of how inhomogeneous media can change
fundamental astrophysical processes and affect the evolution of
different astronomical environments.

Understanding clumpy flow dynamics poses significant scientific
challenges. The vast majority of theoretical treatments of
astrophysical fluid flows have only considered smooth distributions of
gas.  A number of pioneering studies by Dyson, Hartquist and
collaborators have attempted to understand the role of embedded
inhomogeneities via (primarily) analytical methods (e.g., see
\citep{HD86, HD88}). One critical feature of these pioneering works
was treatment of clumps as unresolved sources of mass. This ``mass
loading'' of flows via hydrodynamic and diffusive ablation was shown
to produce important global changes in the flow pattern such as the
transition of the flow into a transonic regime irrespective of the
initial conditions.  In general, it was shown that interactions of a
flow with inhomogeneities might cause significant changes in the
physical, dynamical, and even chemical state of the system.

Attempts to produce fully resolved numerical studies of clumpy flow
dynamics have been hampered by speed and memory requirements of
computers.  In general, detailed studies have focused on interactions
of a {\it single clump} with a global flow \citep*{KMC} (hereafter KMC),
\citep*{Jun99, Lim99, MacLow94}. The advent of adaptive mesh refinement (AMR)
computational technologies has allowed resolved multiple clump systems
to begin to be studied. In \citep*{Pol} (hereafter PFB) a numerical
study of shocks overtaking multiple clumps was completed attempting to
articulate the basic physical processes involved as well as
differentiating clump parameter regimes. In that paper two regimes
were described in which the neighboring clumps either did, or did not,
interact as they were overtaken and destroyed by the flow.  In
addition, it was demonstrated that mixing of clump and ambient
materials was affected by the distribution of clumps.

In spite of the computational advances which made the PFB study
possible an understanding of the full 3-D dynamics of clumpy flows
including the effect of microphysical processes is still not
available. The problem is sufficiently complex that numerical methods
can not be expected to fully articulate answers to the problem. Thus,
additional investigative methods are warranted.  In this paper we
describe the design of a High Energy Density (HED) laboratory
experiment to study shock propagation in clumpy flows. The advent of
HED laboratory methods is a new development in the study of
astrophysical phenomena \citep{Remington1,Remington2} and may offer the
opportunity to probe dynamical processes with control parameters that
are not possible in traditional, observational approaches.

As with the numerical investigations of clumpy flows, the literature
currently contains only experiments examining the interaction of a
shock with a single clump. In \citep{Klein} the NOVA laser was used to
drive a strong shock ($M \approx 10$) into a low-density plastic
target with a single embedded copper microsphere. The morphology and
evolution of the shocked ``cloud'' as well as the trajectory of the
shock were tracked via radiography. The experiment was able to follow
the shock-cloud interaction for a number of dynamical or ``cloud
crushing'' times. These results were compared with 2.5 and 3-D
simulations. The 3-D results used AMR methods that allowed significant
details in the evolution of the shocked cloud to be determined.
\citet{Klein} observed a flattening of the shocked cloud as well as
the appearance of hollow interior. The hollowing was attributed to a
breakup of the vortex ring that forms as the shock traverses the
cloud. \citet{Robey} have reported further studies of this vortex ring
breakup. \citet{KangA} also carried out laboratory experiments of a
shock interacting with a dense sphere embedded in foam. In these
experiments a so-called ``complex shock'' (forward, reverse shock
waves, and the intermediate contact discontinuity) was formed when a
supersonic flow impacted low-density matter.  The flattening of the
cloud was again observed, as was the vortex ring on the downstream
side of the cloud. Using 1 and 2.5-D simulations \citet{KangB} tracked
the evolution of the complex shock as well as the disruption of the
cloud.

In the present paper we describe a design for an experiment that
involves multiple dense clumps interacting with a strong shock wave.
By its nature the clump problem is complex and care must be taken to
create experiments that focus on specific questions lest one ends up
with results too complex to analyze. Issues such as the role of
interactions among clumps, the effects of multiple clumps on mixing
and turbulence, and the nature of mass loading due to clumps could all
be investigated with the correct design but it is unlikely that they
can all be investigated with a single experiment. Thus, in this paper,
we consider a design which is meant to study the global question of
how a clumpy medium affects the dynamics of shock propagation.
Specifically we focus on how a highly clumped medium alters the global
shock propagation speed. This question is relevant to astrophysical
flows, such as shocks traversing giant molecular clouds in which many
cloud cores may exist or the progress of a supernova blast wave
through a clumped wind ejected from the star during a previous epoch.

In section 2 we describe the experimental set-up. In section 3 we
present simulations of the experimental set-up comparing targets with
and without clumps, as well as the targets that have a smooth
distribution of mass whose average density is equal to that in the
clumped target. These results are of interest in their own right
independent of this particular experiment. We note, in particular,
that clumps and foams may be similar enough in principle for our
simulations to bear on general issues related to inertial confinement
fusion (ICF) and laboratory astrophysics experiments. In the final
section we summarize and discuss our results giving prescriptions for
the set-up of these experiments on either intense laser or
pulsed-power experimental testbeds.

\section{EXPERIMENTAL DESIGN}

Several goals for an experiment that is to explore the propagation of
shocks through clumpy media can be identified. 1) The fraction of the
volume occupied by the clumps should be realistic; we chose 5$\%$. 2)
The ratio of density in the clumps to density in the interclump medium
should be realistic. Our choice was 40 to 1. This is on the low end of
observed values; we chose it to maximize the heating of the clumps and
for other reasons discussed below. 3) The shock wave in the clump region
should be reasonably steady and enduring. Specifically, it should be
sustained long enough so that it interacts with many clumps without
any substantial change in the properties of the shock or the
post-shock flow. 4) The experiment should allow comparison of the
clumpy case with alternative cases, in which the shock wave propagates
through media having either a density equal to the average density of
the clumpy medium or a density equal to the interclump density. 5) The
shock wave should be as strong as possible, in order to maximize the
heating of the clumps. Ideally, the clumps should be ionized so that
they can be accurately treated by an ideal-gas equation-of-state with
a polytropic index $\gamma = \sfrac{5}{3}$. 6) The experiment should
be diagnosable using available techniques.

In seeking to meet these goals, we have developed the experimental
design shown in Figure~\ref{target}. The energy source for these
experiments is a pulse power device known as a Z pinch, specifically
the ``Z Machine'' operated by Sandia National Laboratories
\citep{Matzen}. A Z pinch can implode an array of W wires at high
velocity, so that an intense x-ray pulse is produced when the wires
collide and their kinetic energy is thermalized.  Among existing
high-energy-density research facilities, Z can deliver the most energy
to a target.  This is essential for an experiment that requires a
(comparatively) large volume; the present experiment involving many
clumps is a good example.  Through the use of a surrounding, high-Z
container (a hohlraum) to help contain the x-rays, it is feasible to
irradiate up to 4 targets per implosion with an x-ray pulse whose
spectrum is reasonably approximated as a blackbody spectrum with a
temperature of 140 eV. The full-width half-maximum (FWHM) of this
x-ray pulse is 8 ns.

The radiation-hydrodynamic computer code HYADES \citep{Larsen} was
used to evaluate our design options. HYADES is a single-fluid,
Lagrangian code in which the material composition can be different
from cell to cell and in which the electron and ion temperatures
evolve independently.  The electron heat transport is by flux-limited
diffusion, but this was not important here.  The version of the code
used here employs a single-temperature (``greybody'') radiation field
with flux-limited, diffusive radiation transport.  This was only
important in the initial delivery of energy to the target.  The code
was run with SESAME equation-of-state tables.  For our purposes here,
what mattered in the radiation hydrodynamics was to deliver the
correct amount of energy to the initial layer in the target.
Accordingly, we adjusted the radiation temperature to obtain the
correct ablation pressure for the measured radiation temperature,
based on well-confirmed scaling relations \citep{Lindl}.  We also
compared the behavior of targets simulated using a measured radiation
pulse, which includes an extended, low-temperature foot at the start
of the pulse, with the behavior using an approximate pulse of constant
temperature and 8 ns duration. These were very similar (the energy
coupling is dominant). Therefore, a simpler 8 ns pulse was used for
our scaling studies that developed the specific target properties.

The available x-ray pulse cannot be used to directly drive the desired
shock wave. It is too brief, and one needs to absorb it by
solid-density matter before driving the shock through a lower density
material.  The initial challenge in the design is thus to transform
the x-ray energy into hydrodynamic energy, which eventually will be
used to drive the desired steady shock.  The most efficient way to
extract energy from a radiation source is to allow it to accelerate a
thin layer of material over some distance, while ablating some
fraction of the initial layer \citep{Ripin}. The initial layer must be
massive enough to provide the necessary momentum to the additional
mass encountered later during the experiment. It also must be thick
enough that instabilities at the ablation surface do not disrupt it.
Scaling studies showed that a 125 $\mu$m thick layer of polystyrene,
at a density of 1 g/cc, worked well for this purpose. The range of
optimum thickness is not narrow; one would obtain comparable results
from thinner or thicker layers.  The initial layer is accelerated
across a 500 $\mu$m vacuum gap.  The size of the gap was optimized so
that the plastic layer collides with the next layer in the target (the
C foam) just at the end of the 8 ns drive pulse. At this time, it has
been accelerated to a velocity of about 70 km/s.

The second design challenge is to convert the energy initially
delivered by the 8 ns x-ray pulse into a form that can drive a shock
for tens of ns, which turns out to be necessary for reasons discussed
below.  To accomplish this, we let the accelerated plastic layer
impact a C foam layer of density 200 mg/cc and thickness 1.5 mm.  Our
goal here is to let a blast wave develop in the C foam, in which an
abrupt shock is followed by a gradual deceleration over a significant
distance.  The areal mass density of the C foam is several times that
of the plastic that remains when the foam is impacted, so that this
blast-wave becomes approximately a (planar) Sedov-Taylor wave,
decelerating slowly, accumulating mass, and growing spatially.  Our
goal is to use the rarefaction of the leading edge of this blast wave
into the clumpy medium, to drive the enduring shock that we intend to
produce. Here again, we used scaling studies to set the foam density
and thickness. The C foam must be dense enough that its leading edge
can drive the shock we require through the clumpy medium, yet not so
dense that the shocked foam becomes too cool and the rarefaction
becomes too slow.  The changes in behavior were seen to be gradual, so
that the specific parameters of the target represent specific
conditions within a broad range of reasonable choices.

Figure~\ref{1Devolution} shows the evolution of the blast wave as it
reaches the end of the C foam. By these times the initial plastic
layer, seen as the dense feature on the left, has decelerated nearly
to rest.  The local modulations in density and velocity are acoustic
and are artifacts of the zoning used in the simulation. The first
profile shows the blast wave near the end of the C foam. One sees the
characteristic abrupt shock and gradual deceleration. One can also see
that the structure of the blast wave extends over several hundreds of
$\mu$m. This enables it to deliver its energy for several tens of ns,
with characteristic velocities of tens of km/s. When the blast wave
reaches the interclump medium (solid profile), its velocity is $\sim$
33 \kms. At that time, as seen in the other profiles, a rarefaction
wave forms. This drives a faster, reasonably steady shock into the
interclump medium, as is expected for a centered rarefaction
\citep{Zeldovich}\footnote{The density spike at the leading edge of
the rarefaction, which is not important for the overall hydrodynamic
evolution, is an artifact of the equation-of-state tables. It would
not actually be present for the ionized material present here.}. The
rarefaction propagates backward into the C foam as well. However, no
subsequent reflected shock is produced during the experiment because
there are no density structures in the blast wave.

The properties of the clumpy medium were chosen based on several
considerations. The interclump medium must be low in density to allow
the rarefaction of the blast wave to expand at a high velocity and to
allow diagnostic x-rays to penetrate distances of several mm for
radiography.  Subject to these constraints, higher values of this
density produce larger ram pressures and more heating of the clumps.
We chose 25 mg/cc divinyl benzene (DVB) foam for this material. This
produced the minimum realistic ratio of 40 to 1 between the interclump
medium and the plastic clumps at 1 g/cc. The value of 1 g/cc was the
lowest density for clumps that could be implemented, leading to the
maximum clump heating. The clumps were chosen to be two-dimensional
rods since, on one hand, this was easier to diagnose and simpler to
model and, on the other, since an approach to the manufacturing method
of such a system could be identified. The choice of a 5$\%$ volume
fraction for the clumps implied a relation between the radius of the
clumps and the clump density. The choice of a clump size was subject
to the competition between the goals of being able to make them and to
diagnose them, which favored large clumps, and the goals of maximizing
their temperature and of having many clumps in the experiment, which
favored small clumps. The size of the clumps was chosen to be 50
$\mu$m, corresponding to an average interclump spacing of 200 $\mu$m,
a clump density of 25 per square mm, with a total of 200 clumps in a 4
mm wide, 2 mm thick clumpy layer. The location of the clumps in the
clumpy medium was chosen to be random, as is the case in actual
clouds. This was accomplished by using an algorithm that employed a
random number generator to specify these locations.

By using a DVB foam with a density of 73 mg/cc in place of the clumpy
medium, one could observe the propagation of the shock wave through a
medium of uniform density (at least on scales larger than the
$\mu$m-cell size within the foam material).  By using 25 mg/cc DVB
foam without clumps, one could observe the propagation through the
uniform interclump medium.

Beyond the clumpy medium, additional components are placed that allow
the emergence of the shock to be detected and ideally also the
subsequent velocity of the interface at the end of the clumpy medium
to be measured. One approach is to place a quartz window, coated with
a thin ($\sim$1 $\mu$m) Al layer, at the end of the foam and to diagnose
the motion of the Aluminum layer interferometrically.

\section{THEORETICAL AND NUMERICAL ANALYSIS}

\subsection{Theoretical Background}

The interaction of a shock wave with a clumpy medium has recently been
analyzed by PFB in systems with different numbers of clumps and
different clump arrangements. They were simulated numerically using an
AMR code. Here we will give a brief summary of the results and will
refer the reader to that paper for further details.

Analytical arguments drawn from examination of the simulations allowed
two regimes of clumpy flows with distinct flow behavior to be
identified. In the ``interacting'' regime, the evolution of individual
clouds was strongly affected by their neighbors. As is well known
(KMC), the behavior of individual shocked clumps in the adiabatic
regime is dominated by compression and subsequent expansion in a
direction perpendicular to shock propagation (flattening). When this
expansion causes neighboring clumps to interact on a timescale shorter
than the time for them to be destroyed by the post-shock flow, the
subsequent evolution is more appropriately described as a larger
merged system that then progresses toward turbulence. In the
non-interacting regime the clumps are so widely separated that one can
describe their evolution up to destruction in terms of a single
shock-clump interaction.  PFB found that clump distributions in the
interacting regime showed more robust mixing between shock and clump
material apparently due to stronger turbulent motions downstream. The
enhanced mixing seen in PFB may have important consequences in
astrophysical systems such as SNe and evolved stellar wind-blown
bubbles where processed elements in the clumps will be disbursed
through the ISM. In the noninteracting regime clumps will evolve
independently until they are destroyed by the shock and the post-shock
flow.

The concept of interacting and non-interacting regimes can be made
more quantitative. For an external shock velocity $v_S$, a clump
radius $a_0$, and clump to ambient density ratio $\chi = \rho_c
/\rho_a$ the key timescales are the shock crossing timescale $t_{SC}$,
clump crushing timescale $t_{CC}$, and clump destruction timescale
$t_{CD}$,
\begin{eqnarray}
t_{SC} & \equiv & \frac{2a_0}{v_S},\\
t_{CC} & = & \Big(\frac{\chi}{F_{c1}F_{st}}\Big)^{\sfrac{1}{2}}t_{SC},\\
t_{CD} & = & \alpha t_{SC},
\label{timesc}
\end{eqnarray}
where $F_{c1} \approx 1.3$ and $F_{st} \approx 2.06$ for our
experimental conditions. The first quantity relates the stagnation
pressure with the pressure behind the internal forward shock in the
clump, while the second relates the external postshock pressure far
upstream with the stagnation pressure at the cloud stagnation point.
The parameter $\alpha$ is determined from the simulations and is
typically 2. Therefore, systems with a clump density ratio of $\chi =
40$ and a strong external shock (Mach number in the range $M_S = 5 -
100$) yield $t_{CC} \approx 3.9 t_{SC}$ and $t_{CD} \approx 7.8
t_{SC}$.

A critical separation, $d_{crit}$, perpendicular to the direction of
shock propagation can then be derived and expressed as
\beq
\begin{array}{lll}
d_{crit} \! \! & \! \! = \! \! & \! \! \displaystyle
2\big(a_{0}+v_{exp}(t_{CD}-t_{CC})\big) \\ 
\! \! & \! \! = \! \! & \! \! \displaystyle 2a_{0}\Bigg\{
\frac{t_{CD}-t_{CC}}{t_{SC}} \Bigg(\frac{F_{c1}F_{st}}{\chi}\Bigg)^{\sfrac{1}{2}}
\Bigg(\frac{3\gamma(\gamma - 1)}{\gamma + 1}\Bigg)^{\sfrac{1}{2}}+1\Bigg\},
\end{array}
\label{dcrit}
\eeq where $\gamma$ is the adiabatic index of the constituent gas and
$v_{exp}$ is the lateral expansion velocity of the clump equal to the
clump internal sound speed. When clumps are initially separated by a
distance $d > d_{crit}$ they will be destroyed before they interact. A
similar quantity $L_{CD}$, the cloud destruction length, can be
defined for the direction parallel to the direction of shock
propagation. Expressions for $L_{CD}$ are somewhat cumbersome, relying
on a description of the acceleration of the clumps after the passage
of the shock. Readers may find the relevant expressions in PFB.

To summarize, inhomogeneous flows will be in the interacting regime
when initial clump distributions have average separations between
clumps normal to the flow $d$ such that $d < d_{crit}$ and along the
flow $L < L_{CD}$. In what follows we use these results in describing
simulations exploring the experimental design described in the
previous section.

\subsection{Numerical Setup and Method}

A series of numerical simulations based on the experimental design
presented in Section 2 were performed. The simulations used the
Adaptive Mesh Refinement (AMR) hydrodynamics code AMRCLAW
\citep{Berger98}. AMR methods allow high resolution to be applied
dynamically where needed in a calculation.  Such methods are critical
in studies of high Mach number clumpy flows as it would be impossible
to properly resolve both the details of individual shock-clump
interactions and the global flow with fixed grid methods. Description
of the code and its application to the clumpy flow problem can be
found in PFB.

Our clumpy flow simulations were performed in 2-D Cartesian geometry
and are therefore slab symmetric. The simulations were initialized
with a blast wave of properties described in Section 2 and
corresponding to the solid profile in Figure~\ref{1Devolution},
propagating into an array of 200 clumps distributed randomly. The
properties of the clumps and surrounding media were also the same as
described in Section 2. The particular distribution of clumps (each
with $a_0 = 25$ $\mu$m and $t_{SC} = 0.96$ ns) used in the simulations
had the following properties: 
\beq
\begin{array}{lllll}
d_{crit} & = & 4.26 \ a_0   & =       & 106.5 \ \mu m, \\
L_{CD}   & = & 3.54 \ a_0   & =       & 88.5  \ \mu m.
\end{array}
\label{params}
\eeq 
Average clump separations in the experiment, calculated according to the
expressions (39) and (40) of PFB, are
\beq
\begin{array}{lllll}
\langle \Delta x \rangle  & = & 5.35  \ \mu m & \approx & 6.05\% \ L_{CD}, \\
\langle \Delta y \rangle  & = & 10.24 \ \mu m & \approx & 15.88\% \ d_{crit}.
\end{array}
\label{dxdy1}
\eeq 
Thus, the system designed for the experiments can be described as strongly
interacting with both global and local evolution strongly affected by
clump merging prior to breakup.

The details of our simulations were as follows. We used a 3 level
system of refinement corresponding to a maximum equivalent resolution
of $3264 \times 2560$ zones. Thus, at maximum resolution each clump
radius $a_0$ was resolved with at least 16 zones. That resolution is
significantly smaller than the resolution that would correspond to the
converged regime (e.g., see \citep{KMC,Pol}). However, on one hand,
such converged resolution was unfeasible with given computational
resources. On the other hand, achieved resolution was sufficient to
capture global properties of shock propagation in the clumpy medium
which was the primary focus of investigation. The computational domain
had dimensions $160 a_0 \times 204 a_0$. The shock entered the domain
from the left. All four boundaries of the domain had outflow boundary
conditions. In general, however, high density material, corresponding
to the target walls and included in the computational domain, kept the
flow from reaching the side boundaries for a rather large part of the
simulation. Similarly, while the far side of the computational domain
was an open boundary, the high density backplate kept the flow from
exiting the grid during the simulation.

Three simulations were run corresponding to the likely program of
experimental shots. Run 1 contained no clumps, Run 2 contained the
clump distribution described above, and Run 3 contained no clumps but
had a smooth background whose density was equal to the average density
in the clump region of the clump run. All three runs lasted well past
the moment of contact of the global shock with the backplate, namely
the Run 1 was run for $t = 76$ ns or $\approx 79 t_{SC}$, Run 2 was
run for $t = 138.2$ ns or $\approx 144 t_{SC}$, and Run 3 was run for
$t = 70$ ns or $\approx 73 t_{SC}$\footnote{Hereafter in our
simulations time is given from the moment when the blast wave enters
the low-density interclump medium.}.

\subsection{Results}

Figure~\ref{frame} shows a snapshot of the clumpy simulation early in
its evolution after 15.89 ns (16.55 $t_{SC}$) and is a synthetic
Schlieren representation of the logarithmic density in the flow. A
number of points are worth noting. By this time the first ``row'' of
clumps has already been destroyed and the material from these clumps
has been accelerated downstream. The effect of this first line of
clumps on the downstream neighbors can be clearly seen in this figure.
The second ``row'' of clumps is disrupted both by the passing of the
global shock as well as by the debris from the upstream clumps. Thus,
the kinetic energy in the flow $F_k$ interacting with these clumps is
enhanced above that for single clumps as 
\beq 
F_k = \frac{1}{2}\rho_{PS}v^2_{PS}+F_{C}= \frac{2\rho_{0}v^2_{S}}
{(\gamma-1)(\gamma+1)}+\frac{1}{2}\langle\rho_{CA}\rangle v^2_{C}.
\label{kinflux}
\eeq 
Here $\rho_{PS}$ and $v_{PS}$ are the post-shock density and
velocity respectively, $\rho_{0}$ is the unshocked interclump density,
and $v_{S}$ is the global shock velocity. The first term corresponds
to the undisturbed post-shock flow. The second term encompasses the
accelerated clump material and it depends on its instantaneous
velocity $v_{C}$ (expressions for it can be found in PFB) as well as
the details of its dispersion, i.e.  $\langle \rho_{CA} \rangle$. As
was shown in PFB, such interactions greatly enhance mixing, and more
importantly for our purposes, rob the shock of energy by converting
bulk flow into a turbulent cascade of vorticity.

In Figure~\ref{frame} notice also the presence of the individual bow
shocks surrounding each clump. As the incident shock progresses, these
individual shocks merge into a single structure.  This merged
structure becomes normal to the direction of the flow eventually
taking the form of a single reverse shock. Thus, we see a clear
transition from a flow pattern in which the heterogeneity dominates
initially to one in which a global flow dynamics emerges. It is
worthwhile noting that considerable acoustic structure can be seen
behind (to the left) of the global reverse shock as information about
newly shocked clumps propagates upstream.

Finally, notice that while the global shock has numerous corrugations
in it due to the presence of the clumps we do not see any large scale
disturbances. This trend continues for the entire evolution of the
simulation. On average, the shock appears to ``anneal'' itself as it
passes through the clump region. The global structure of the shock 
upon passing through the clumpy region and the effect of ``annealing''
can be studied in the experiment via the diagnostics of the interaction
of the shock with the Al layer on the backplate.

The global behavior of the incident shock is illustrated in
Figure~\ref{shockpos} that shows the progress of the shock as a
function of time for all three simulations. For the clumpy simulation
this figure was made by constructing an average position for the shock
from different samplings along strips in y-direction. The process was
performed by locating the global shock position. An estimated error is
of about the cell size at the level with the highest resolution, which
is about 1.6 $\mu$m. This figure demonstrates how shock propagation
will differ in a clumpy flow from that in a smooth one, even when the
smooth flow has the same average properties. Initially, shock front
velocity in the clumpy case is virtually identical to that when the
clumps are absent. However, as the shock propagates further through
the clumpy region, causing more clumps to be destroyed, its velocity
gradually starts to decrease. As the whole clumpy system breaks up and
homogenizes, the global shock velocity would eventually tend toward
the values that correspond to the case of a uniform averaged density
medium. The time for the clumpy system to change its behaviour from
the one similar to the case with no clumps to the one similar to the
case with average density is on the order of the system destruction
time $t_{SD}$. Definition of $t_{SD}$ is given in section 3.1.2 of
PFB. It should be noted, however, that from the experimental point of
view it was unfeasible to provide extended region behind the clump
section of the target to follow clump system evolution after its
break-up for some period of time. Since in our simulations we tried to
reproduce the target design as closely as possible, we were not able
to observe the shock front velocity evolution beyond the point of its
emergence from the clump system and contact with the backplate.

Average values for the shock velocities derived from
Figure~\ref{shockpos} are: $\langle v_{s1} \rangle \sim 57.1$ \kms;
$\langle v_{s2} \rangle \sim 51.95$ \kms; $\langle v_{s3} \rangle \sim
44.94$ \kms. Thus the shock in the clumpy case propagates 15.6$\%$
faster than in the situation with the same average density but no
inhomogeneities. This is an important point that may have significant
consequences for astrophysical flows as we discuss in the final
section. From the design standpoint, however, note that the time
resolution associated with the diagnostic modes of shock detection to
be used in the experiment is about 100 ps, whereas the difference
between the shock arrival time at the backplate for those two cases is
about 6 ns. Thus, data taken from the shots should be capable of
distinguishing between the various cases - an important point for the
development of a successful test bed for studying clumpy flows.

Figures~\ref{backlighter1} and \ref{backlighter2} show synthetic X-ray
backlighter images of the clumpy simulation at 15.89 ns.
Figure~\ref{backlighter1} is an image using 5 keV X-rays with a
maximum optical density on the film of 2.0, and
Figure~\ref{backlighter2} is an image using 10 keV X-rays with the
same maximum film density. The images were calculated using cold
material x-ray attenuation coefficients and a linear log(exposure) to
optical density relation.

The 5 keV X-ray radiograph of Figure~\ref{backlighter1} shows the
structure of the global shock wave as it propagates through the medium
in which the clumps are embedded. The shock front and subsequent shock
waves from the clumps are clearly visible. The effect of the shock
wave on the clumps is shown by the higher energy X-ray radiograph of
Figure~\ref{backlighter2}. The collapse of individual clumps as the
shock wave interacts with them is apparent.

\section{DISCUSSION AND CONCLUSIONS}

In this paper we have presented a design for an experiment to explore
the evolution of a strong blast wave propagating through a clumpy
medium.  These experiments are relevant to astrophysical shocks
propagating through a variety of inhomogeneous environments including
young stellar objects, supernova remnants, planetary nebulae, and AGN.
Our experiment begins with a shock wave created by the acceleration of
a slab of material via energy deposition from a pulsed power source.
The blast wave propagates first through a smooth region of low density
foam followed by a region in which plastic rods are embedded in a foam
background.  We have described the nature of the blast wave and the
properties of the foams that constitute the ``ambient medium'', as
well as the rods that constitute the clumps. We note that, while these
experiments are described in the context of the Z-pinch pulsed power
machine, our design could be adopted to other high energy density
devices such as high intensity lasers \citep{Omega,NIF}.

Numerical simulations based on the experimental design confirm that
the experiments should be capable of exploring differences between
shocks propagating though: 1) smooth media; 2) clumpy media; 3) smooth
media whose properties correspond to the average properties of clumpy
regions. Our simulations show that shocks in clumpy media move more
rapidly than those in the smooth media with averaged properties. The
shock speed in the former case is 7.01\kms, or 15.6$\%$, higher than
in the latter for the experimental conditions. This difference is
sufficiently large to be resolved by existing diagnostics. Thus, the
experiments should be able to explore issues surrounding energy
deposition in the evolution of clumpy flows.

Our simulations have already revealed behavior that may bear on the
evolution of clumpy astrophysical flows. Our results indicate that the
use of average properties of inhomogeneous regions in calculations of
critical timescales may be inappropriate and lead to incorrect
conclusions. Consider, for example, the propagation of a strong shock
impinging on a molecular cloud of size L which contains many dense
cores. Considering that the shock compression of the cores may trigger
gravitational collapse, one would be interested in comparing the cloud
crossing timescale for the shock to propagate across the cloud to the
time associated with the collapse of an individual core.
Determination of the relative size of these two timescales would
establish the environments in which star formation would occur. For
example, when $t_{cross} << t_{col}$ then one can ignore any ongoing
interaction between the shock and the collapsing core.  Our results,
however, indicate that care is needed in determination of $t_{cross}$
since $t_{cross} = L/v_{S}$ where $v_{S}$ is the shock speed in the
cloud. Since $v_{S}$ is a function of conditions in the cloud such as
the density $\rho_c$, our results indicate that replacing the
spatially inhomogeneous $\rho_c(x,y,z)$ (appropriate to a clumpy
medium) with the average $\langle \rho_c \rangle$ can lead to an
overestimation of the $t_{cross}$. While these issues require more
study, the simulations presented in this paper (along with those
presented in PFB) indicate that global dynamics of clumpy flows may
not be easily captured by simple smoothing through the use of spatial
averages.

These results may also have implications for the behavior of foam
targets themselves. Some foams are composed of many small bubbles, so
that a shock wave will propagate through a random sequence of bubble
walls, causing them to expand and creating many small local shocks and
rarefactions. Other foams have the morphology of a pile of straw with
long rods separated by spaces. This case has more resemblance to that
studied here, as the shock will propagate around the rods, which will
subsequently be destroyed. However, the vacuum between the rods
implies that in this case also local rarefactions will play an
important role (unless preheat has caused the release of gas
throughout the foam). In both these cases, the actual equation of
state of the foam is complex and depends on the history of the
material, at least when preheat and shocks are present. In addition,
the post-shock state of the foam is likely to include turbulent
motions that take up a non-negligible energy fraction, which implies
that the EOS of the shocked foam may differ from that of an ordinary
plasma for some period of time. Studies similar to those reported
here, but in better optimized morphologies, could contribute to better
understanding of the detailed behavior of these systems.

\acknowledgements This work benefited greatly from discussions with
Tom Gardiner (University of Maryland), Eric Blackman (University of
Rochester), and Marcus Knudson, Clint Hall, Michael Cuneo, and Tom
Mehlhorn (Sandia National Laboratories). Support to AYP and AF was
provided at the University of Rochester by NSF grant AST-9702484 and
AST-0098442, NASA grant NAG5-8428, and the Laboratory Astrophysics
program at the Laboratory for Laser Energetics. Support for work by
the University of Michigan was provided by the Sandia National
Laboratories and by the High-Energy-Density Sciences Grants program at
the USDOE. We also acknowledge work on the foam technology and
production by Diana Schroen of Schaefer Corporation.

The most recent results and animations of the numerical experiments,
described above, as well as the ones not mentioned in the current
paper, can be found at \url{www.pas.rochester.edu$/^{\sim}$wma}.


\clearpage

\begin{figure}
\caption{The target and its components, described in the text.
\bf{Please, see additional file figure1.eps for the figure.}
\label{target}}
\end{figure}

\begin{figure}
\caption{Evolution of the density and velocity as the blast wave enters
the low-density interclump medium. The profiles are labeled with the
time in ns. \bf{Please, see additional file figure2.eps for the figure.}
\label{1Devolution}}
\end{figure}

\begin{figure}
\caption{Synthetic Schlieren image of the density logarithm from the simulation of 
the experimental design with clump region. The shock propagates toward
the right into the region of plastic rods. Image taken 15.89 ns after the blast
wave entered the low-density interclump foam. Note the bow-shocks forming around 
the individual clumps and the destruction of the first row of rods. Position
values correspond to the ones in Figures~\ref{target} and \ref{1Devolution}.
\bf{Please, see additional file figure3.eps for the figure.}
\label{frame}}
\end{figure}

\begin{figure}
\caption{Plot of shock position vs. time for three simulations: 1) Run 1: no clumps, 
background density only; 2) Run 2: clumps; 3) Run 3: no clumps but with background
density equal to the average density in the target with clumps. The
two horizontal lines indicate the extent of the clumpy region in the
simulations and the target. \bf{Please, see additional file figure4.eps for the figure.}
\label{shockpos}}
\end{figure}

\begin{figure}
\caption{Synthetic X-ray back lighter image of the clumpy simulation at 15.89 ns 
for the 5 keV X-rays. \bf{Please, see additional file figure5.eps for the figure.}
\label{backlighter1}}
\end{figure}

\begin{figure}
\caption{Synthetic X-ray back lighter image of the clumpy simulation at 15.89 ns for
the 10 keV X-rays. \bf{Please, see additional file figure6.eps for the figure.}
\label{backlighter2}}
\end{figure}

\end{document}